\documentclass{article}
\usepackage{amsthm,amsmath,amssymb,amscd}

\makeatletter

\def\@aabuffer{}
\def\author #1{\expandafter\def\expandafter\@aabuffer\expandafter
{\@aabuffer \small\rm      #1\relax \par}}
\def\address#1{\expandafter\def\expandafter\@aabuffer\expandafter
{\@aabuffer \small\it #1\relax \par\vspace{1em}}}

\def\maketitle{
\begin{center}
   {\bf \@title \par}       
   \vskip 2em                      
   \@aabuffer\relax
\end{center} \par
\gdef\@aabuffer{}
}

\def\abstracts#1{
\begin{center}
{\begin{minipage}{4.2truein}
                 \footnotesize
                 \parindent=0pt #1\par
                 \end{minipage}}\end{center}
                 \vskip 2em \par}

\fussy
\def\section{\@startsection {section}{1}{\z@}{-3.5ex plus -1ex minus 
    -.2ex}{2.3ex plus .2ex}{\bf }}
\def\subsection{\@startsection{subsection}{2}{\z@}{-3.25ex plus -1ex 
    minus -.2ex}{1.5ex plus .2ex}{\it }}

\def\@makefnmark{{$^{\@thefnmark}$}}

\renewenvironment{thebibliography}[1]
        {\begin{list}{\arabic{enumi}.}
        {\usecounter{enumi}\setlength{\parsep}{0pt}
         \setlength{\itemsep}{0pt} 
         \settowidth
        {\labelwidth}{#1.}\sloppy}}{\end{list}}

\newcounter{arabiclistc}

\def\@citex[#1]#2{\if@filesw\immediate\write\@auxout
        {\string\citation{#2}}\fi
\def\@citea{}\@cite{\@for\@citeb:=#2\do
        {\@citea\def\@citea{,}\@ifundefined
        {b@\@citeb}{{\bf ?}\@warning
        {Citation `\@citeb' on page \thepage \space undefined}}
        {\csname b@\@citeb\endcsname}}}{#1}}

\newif\if@cghi
\def\cite{\@cghitrue\@ifnextchar [{\@tempswatrue
        \@citex}{\@tempswafalse\@citex[]}}
\def\citelow{\@cghifalse\@ifnextchar [{\@tempswatrue
        \@citex}{\@tempswafalse\@citex[]}}
\def\@cite#1#2{{$\!^{#1}$\if@tempswa\typeout
        {IJCGA warning: optional citation argument 
        ignored: `#2'} \fi}}

\def\baselinestretch{1.0}
\ifx\selectfont\undefined
\@normalsize\else\let\glb@currsize=\relax\selectfont
\fi

\ifx\selectfont\undefined
\def\@singlespacing{%
\def\baselinestretch{1}\ifx\@currsize\normalsize\@normalsize\else
\@currsize\fi%
}
\else
\def\@singlespacing{\def\baselinestretch{1}\let\glb@currsize=
\relax\selectfont}
\fi

\long\def\@makecaption#1#2{
   \vskip 10pt 
   \setbox\@tempboxa\hbox{\footnotesize #1: #2}
   \ifdim \wd\@tempboxa >\hsize   
       \leftskip 0pt plus 1fil 
       \rightskip 0pt plus -1fil 
       \parfillskip 0pt plus 2fil 
       \footnotesize #1: #2\par   
     \else                        
       \hbox to\hsize{\hfil\box\@tempboxa\hfil}  
   \fi}

\makeatother

\DeclareMathOperator{\Rea}{Re}
\DeclareMathOperator{\Ima}{Im}
\DeclareMathOperator{\Hom}{Hom}
\DeclareMathOperator{\Li}{Li}
\DeclareMathOperator{\Lir}{\mathbb{L}i}
\DeclareMathOperator{\Lic}{{\cal L}i}
\DeclareMathOperator{\Pic}{Pic}
\numberwithin{equation}{section}

\title{ STRING DUALITY AND ENUMERATION OF CURVES BY JACOBI FORMS}

\author{TOSHIYA KAWAI}

\address{Research Institute for Mathematical Sciences,\\ Kyoto
  University, Kyoto 606--8502, Japan}

\date{}

\begin{document}

\newcommand{\CC}{\mathbb{C}}
\newcommand{\DD}{\mathbb{D}}
\newcommand{\RR}{\mathbb{R}}
\newcommand{\QQ}{\mathbb{Q}}
\newcommand{\ZZ}{\mathbb{Z}}
\newcommand{\HH}{\mathbb{H}}
\newcommand{\NN}{\mathbb{N}}
\newcommand{\PP}{\mathbb{P}}
\newcommand{\abs}[1]{\lvert#1\rvert}
\newcommand{\inv}[1]{\frac{1}{#1}}

\maketitle \abstracts{ For a Calabi-Yau threefold admitting both a
  $K3$ fibration and an elliptic fibration (with some extra
  conditions) we discuss candidate asymptotic expressions of the genus
  $0$ and $1$ Gromov-Witten potentials in the limit (possibly
  corresponding to the perturbative regime of a heterotic string)
  where the area of the base of the $K3$ fibration is very large.  The
  expressions are constructed by lifting procedures using nearly
  holomorphic Weyl-invariant Jacobi forms. The method we use is
  similar to the one introduced by Borcherds for the constructions of
  automorphic forms on type IV domains as infinite products and
  employs in an essential way the elliptic polylogarithms of Beilinson
  and Levin.  In particular, if we take a further limit where the base
  of the elliptic fibration decompactifies, the Gromov-Witten
  potentials are expressed simply by these elliptic polylogarithms.
  The theta correspondence considered by Harvey and Moore which they
  used to extract the expression for the perturbative prepotential is
  closely related to the Eisenstein-Kronecker double series and hence
  the real versions of elliptic polylogarithms introduced by Zagier.}

\section{Introduction}

Topological sigma models with target Calabi-Yau threefolds have been
extensively investigated in recent years.  The study of the so-called
A-model [Wit] starts with the following setup.

Let $X$ be a Calabi-Yau threefold. The non-trivial Hodge numbers
$h^{11}$ and $h^{12}$ of $X$ are related to the Euler characteristic
of $X$ by $\chi =2(h^{11}-h^{12})$.  We assume that $H_2(X,\ZZ)$ is
torsion-free.  We identify $H^2(X,\ZZ)$ with $H_4(X,\ZZ)$ and
interchangeably use both viewpoints. Let $D_1,\ldots,D_l$ be the
divisors generating $H^2(X,\ZZ)\cong H_4(X,\ZZ)$ where $l:=h^{11}$.
Introduce the topological invariants
\begin{equation}
  \begin{split}
    \kappa_{ijk}&:=D_i\cdot D_j \cdot D_k\,,\\ 
   \rho_i&:= c_2(X)\cdot D_i\,,
\end{split}
\end{equation}
where $c_2(X)$ is the second Chern class of $X$.  In what follows we
assume that the divisors $D_1,\ldots,D_l$ are nef, {\it i.e.\/}
$D_i\cdot C\ge 0$ $(i=1,\ldots,l)$ for any irreducible algebraic curve
$C$ in $X$.  Hence $\rho_i \ge 0$ $(i=1,\ldots, l)$ by [Mi].

For a fixed complex structure on $X$, the A-model moduli parameters
are conveniently expressed by the complexified K{\" a}hler form
$J_\CC$ belonging to some open subset of the complexified K{\" a}hler
cone $\{J_\CC=B+\sqrt{-1}J \mid B \in H^2(X,\RR)/H^2(X,\ZZ), J\in
\mathcal{K}\}$ where $\mathcal{K}$ is the K{\" a}hler cone of $X$ and
$B$ is associated with the `$B$-field' in the sigma model
language\footnote{For a careful treatment of the A-model moduli space
  see [Mo1,Mo2].}.  Near the large-radius limit $J_\CC$ can be
parametrized as
\begin{equation}
  J_\CC= \sum_{i=1}^{l} \left(\frac{1}{2\pi \sqrt{-1}}\log q_i\right)
  D_i\,, \quad \abs{q_i}\ll 1\,,
\end{equation}
where $\frac{1}{2\pi \sqrt{-1}}\log q_i$ $(i=1,\ldots,l)$ are the
so-called flat coordinates. They are multi-valued and can be shifted
by arbitrary integers.

The  polylogarithm  functions are defined by
\begin{equation}\label{polylog}
  \Li_r(\xi)=\sum_{n=1}^\infty\frac{\xi^n}{n^r},\quad \abs{\xi}<1\,,
  \quad r\in \NN\,.
\end{equation}
The prepotential or the genus $0$ Gromov-Witten potential of $X$ is
known to have the following world-sheet instanton expansion:
\begin{equation}\label{prepo}
  F_0(q_*)=\frac{1}{6}\sum_{i,j,k}\kappa_{ijk} (\log q_i) (\log q_j)
  (\log q_k)-\frac{\chi}{2} \zeta(3) + \sum_{d\in {\cal S}}
  N_0(d)\Li_3(q^d)\,,
\end{equation}
where $\zeta(s)=\sum_{n=1}^\infty n^{-s}$ $(\Rea s>1)$ is the Riemann
zeta function and $q^d:=q_1^{d_1}\cdots q_l^{d_l}$ for
$d=(d_1,\ldots,d_l)\in {\cal S}:={\NN_0}^{l}\smallsetminus \{0\}$. The
(virtual) numbers of rational curves on $X$ are counted by $N_0(d)$'s
which are related to the Gromov-Witten invariants for genus $0$ curves
with at least $3$ marked points [AM,KM].  (The Gromov-Witten
invariants are defined for genus $g$ curves with $n$ marked points
satisfying $2-2g-n<0$.)

On the other hand,
the genus $1$ Gromov-Witten potential of $X$ is believed to have the
following expansion [BCOV]\footnote{Here we are taking an
  unconventional overall normalization and the undetermined constant
  has been set to zero.}:
\begin{equation}
  F_1(q_*)=-\frac{1}{2}\sum_i \rho_i \log q_i+ \sum_{d\in
      {\cal S}}N_{1,0}(d) \Li_1(q^d)\,,
\end{equation}
where 
\begin{equation}
  N_{1,0}(d)=N_0(d)+12 \sum_{\substack{d'\in
      {\cal S}\\ d'\le d}} N_1(d')\,,
\end{equation}
and we have considered $\mathcal{S}$ as a poset by the partial order:
$d'\le d$ $ (d,d' \in \mathcal{S})$ iff $d_i' \mid d_i$ $ ({}^\forall
i)$.  The $N_1(d)$'s predict the numbers of elliptic curves on $X$ and
are related to the Gromov-Witten invariants for genus $1$ curves with
at least $1$ marked points.

After the pioneering work [CdGP] there have been an abundance of works
on the computations of $F_0$ and $F_1$ for various Calabi-Yau
threefolds in which they reduced the tasks to B-model calculations
using mirror symmetry. See for instance [Mo1] and references therein.
However, not much is known about the instanton expansions of the genus
$g$ Gromov-Witten potentials $F_g(q_*)$ when $g \ge 2$.

In this work we are concerned with $F_0$ and $F_1$ when $X$ admits a
$K3$ fibration as well as an elliptic fibration (with some extra
conditions).  $K3$ fibered Calabi-Yau threefolds have been of great
interest in recent developments of string theory, due to their
relevance to heterotic-type IIA string duality conjectures
[KV]. Namely, type IIA string compactified on a $K3$ fibered
Calabi-Yau threefold is a potential candidate for the dual of a
heterotic string compactified on $K3\times T^2$ [KLM, AL, A]. If $D_1$
corresponds to the generic fiber of the $K3$ fibration, the flat
coordinate $\frac{1}{2\pi \sqrt{-1}}\log q_1$ is linearly related to
the dilaton-axion of heterotic string and the limit $q_1\rightarrow 0$
where the area of the base of the $K3$ fibration is very large
corresponds to the perturbative limit of heterotic string. Thus it is
important to understand the behaviors of $F_0$ and $F_1$ in this limit
in order to test duality conjectures and there have already been many
works on the subject, the list of which includes
[KLT,HaM,CCLM,Kaw1,CCL,HeM,C]. The additional assumption about the
existence of an elliptic fibration is also important in string theory
in the context of $F$-theory [V][MV]. It is related to the existence
of the decompactification limit of the $T^2$ factor of the $K3\times
T^2$ on which heterotic string was compactified. We refer to [MV,A]
for more details.

The purpose of this paper is, with the above assumptions on $X$, to
discuss certain expressions which are conjecturally suitable for the
descriptions of $F_0$ and $F_1$ near the limit $q_1\rightarrow 0$. To
give such expressions we use the lifting procedures for some 
 nearly holomorphic Weyl-invariant Jacobi forms similar to the one
introduced by Borcherds [Bo1] in his construction of automorphic forms
on type IV domains as infinite products.  These operations involve the
Hecke operators for Jacobi forms and, more interestingly, the elliptic
polylogarithms of Beilinson and Levin [BeL] in an essential way.  In
particular if we consider the limit where the base of the elliptic
fibration decompactifies, $F_0$ and $F_1$ can be simply expressed in
terms of these elliptic polylogarithms.

In an important paper [HaM], by which the present work is much
influenced, Harvey and Moore introduced a kind of theta correspondence
and extracted the perturbative prepotential of heterotic string from
the computations of some modular integrals.  The relation between the
two approaches, one by lifting and the other by modular integrals, to
investigate the limiting behaviors of $F_0$ and $F_1$ may be partially
explained by the following fact.  In [BeL], they introduced the notion
of the elliptic polylogarithm in their effort to give a motive
theoretic interpretation on the work of Zagier [Z] where the real
versions of elliptic polylogarithms were introduced as the elliptic
extensions of certain real combinations of polylogarithms.  As shown
in [Z] Zagier's elliptic polylogarithms are equal to certain double
series (which we call the Eisenstein-Kronecker double series). These
double series are generalizations of the ones appearing in the
classical Kronecker limit formulas. However, as we will remark in this
paper, Zagier's real combinations of polylogarithms and the
Eisenstein-Kronecker double series naturally appear in the evaluation
of some modular integrals.

The organization of this paper is as follows. In \S 2, we summarize
several properties of (nearly holomorphic) Weyl-invariant Jacobi forms
which will be frequently used in later sections. In \S 3 we consider
the theta correspondence associated with a nearly holomorphic
Weyl-invariant Jacobi form following the approach of [HaM] and discuss
its relation to the Eisenstein-Kronecker double series and the real
version of elliptic polylogarithms [Z]. In \S 4 we discuss the
relation between the Borcherds lifting [Bo1] and the elliptic
polylogarithms introduced in [BeL].  \S 5 is devoted to applications
to the problem described above providing several explicit (albeit
conjectural) examples. There are many things yet to be clarified. We
shall discuss some of them in \S 6.

\bigskip {\noindent \bf Notation.} $\NN$ is the set of positive
integers. $\NN_0$ is the set of non-negative integers. $\mathbf{e}[x]$
stands for $\exp(2\pi \sqrt{-1} x)$.

\section{Weyl-invarinat Jacobi Forms}

 Let $\HH$ be the complex upper half plane $\{\tau\in\CC \mid
\Ima\tau>0\}$.
The normalized
Eisenstein series of weight $k\, (\in 2\NN)$  is defined by
\begin{equation}\label{eisenstein}
  E_k(\tau)=1-\frac{2k}{B_k}\sum_{n=1}^\infty
  \sigma_{k-1}(n)q^n\,,\quad \tau\in \HH\,,
\end{equation}
where $q=\mathbf{e}[\tau]$ and $\sigma_m(n)$ is the sum of the $m^{\rm
  th}$ powers of all the positive divisors of $n$. The Bernoulli
numbers $B_n\ (n\in \NN_0)$ are defined  by
\begin{equation}
  \frac{t}{e^t-1}=\sum_{n=0}^\infty B_n\frac{t^n}{n!}\,.
\end{equation}
Thus, $B_{2n+1}=0$ $(n\ge 1)$ and 
\begin{equation}
  B_0=1,\quad B_1=-\frac{1}{2},\quad B_2=\frac{1}{6}, \quad
  B_4=-\frac{1}{30},\ \ldots
\end{equation}

Let $\mathfrak{g}$ be a simple Lie algebra of rank $s$
with a fixed Cartan subalgebra $\mathfrak{h}$. We identify
$\mathfrak{h}$ and $\mathfrak{h}^*$ using the Killing form $(\ ,\ )
$. We extend $(\ ,\ )$ by $\CC$-linearity.  We normalize the highest
root $\theta$ as $(\theta,\theta)=2$. Let $Q^\vee$ be the coroot
lattice of $\mathfrak{g}$. Then $Q^\vee$ is a positive definite even
integral lattice of rank $s$. Let $P$ be the weight lattice of
$\mathfrak{g}$ so that $P^*=Q^\vee$.  The Weyl group of $\mathfrak{g}$
is denoted by $W$.

 A holomorphic function $\phi_{k,m}:\HH\times
\mathfrak{h}_\CC \rightarrow \CC$ $(k\in \ZZ, m\in \NN)$ is called a
Weyl-invariant Jacobi form of weight $k$ and index $m$ if it satisfies
the following conditions [Wir]:
\begin{enumerate}
\item  For any  $\bigl(\begin{smallmatrix}
    a&b\\c&d
  \end{smallmatrix}\bigr)\in SL_2(\ZZ)$,
  \begin{equation}\label{modular}
 \phi_{k,m}\left(\frac{a\tau+b}{c\tau+d}, \frac{z}{c\tau+d}\right)
  = (c\tau+d)^k {\bf
    e}\left[\frac{mc(z,z)}{2(c\tau+d)}\right]\phi_{k,m}(\tau,z)\,.
\end{equation}

\item  For any $\lambda,\mu\in Q^\vee$,
  \begin{equation}
\phi_{k,m}(\tau,z+\lambda\tau+\mu)={\bf
    e}\left[-m\left(\frac{(\lambda,\lambda)}{2}\tau +
      (\lambda,z)\right) \right]\phi_{k,m}(\tau,z)\,.
\end{equation}

\item  For any $w\in W$,
  \begin{equation}\label{weyl}
  \phi_{k,m}(\tau,w z)=\phi_{k,m}(\tau,z)\,.
  \end{equation}

\item $\phi_{k,m}$
  can be expanded as:
  \begin{equation}\label{expansion}
\phi_{k,m}(\tau,z)=\sum_{\substack{n\in \NN_0\\ \gamma \in P}}
  c(n,\gamma)q^n\zeta^\gamma\,.
\end{equation}

\end{enumerate}
Here   $\zeta^\gamma={\bf e}[(\gamma,z)]$.

  The coefficients $c(n,\gamma)$ depend only on
$mn-\frac{(\gamma,\gamma)}{2}$ and $\gamma\bmod{mQ^\vee}$. They are
zero except for a finite number of $\gamma \in P$ for a fixed $n$. 
It follows from \eqref{modular} that
\begin{equation}\label{reflection}
  \phi_{k,m}(\tau,-z)=(-1)^k\phi_{k,m}(\tau,z)\,.
\end{equation}  
Hence $k$ must be an even integer if $\epsilon\in W$ where $\epsilon$
acts linearly on $\mathfrak{h}_\CC$ as $\epsilon(z)=-z$,
$(z\in\mathfrak{h}_\CC)$.  This is the case if $\mathfrak{g}$ is none
of: $\mathsf{A}_n$ $(n\ge 2)$, $\mathsf{D}_{2n+1}$ $(n \ge 2)$,
$\mathsf{E}_6$.

If we allow $\phi_{k,m}$ to have poles at the cusps so that a finite
number of negative powers of $q$ are allowed in the sum in
\eqref{expansion}, it is called a nearly holomorphic Weyl-invariant
Jacobi form. The vector space of (nearly holomorphic) Weyl-invariant
Jacobi forms of weight $k$ and index $m$ is denoted as
$J_{k,m}(\mathfrak{g})$ ($J_{k,m}^{\rm nh}(\mathfrak{g})$).

Let $\alpha_1^\vee,\ldots,\alpha_s^\vee$ be the simple coroots of
$\mathfrak{g}$ and let $a_0^\vee,\ldots,a_s^\vee$ be the positive
integers such that $a_0^\vee=1$ and $\theta^\vee=\sum_{i=1}^s a_i^\vee
\alpha_i^\vee$ where $\theta^\vee$ is the highest coroot of
$\mathfrak{g}$.  It was shown in [Wir] that if $\mathfrak{g} \ne
\mathsf{E}_8$ the bigraded ring $\oplus_{k,m}J_{k,m}(\mathfrak{g})$ is
a polynomial ring over $\CC[E_4,E_6]$ freely generated by
$\varphi_0,\ldots,\varphi_s$ with the weight and index of $\varphi_i$
given respectively by $-d_i$ and $a_i^\vee$. Here $d_0=0$ and
$d_i=m_i+1$ $(i\ne 0)$ with $m_1,\ldots,m_s$ being the exponents of
$\mathfrak{g}$. See also the earlier works [L,BS,Sai]. For
$\mathfrak{g}=\mathsf{A}_1$, this structure theorem reduces to Th 9.3
in [EZ] (see also [FF]).

Let us fix 
\begin{equation}
  \phi_{k,m}(\tau,z)=\sum\limits_{n,\gamma}
  c(n,\gamma)q^n\zeta^\gamma \in J_{k,m}^{\rm nh}(\mathfrak{g})\,,
\end{equation} 
for an even integer $k$.  We now list up miscellaneous identities
satisfied by the coefficients $c(n,\gamma)$ for later use.  The set
$P_0:=\{\gamma\in P \mid c(0,\gamma) \ne 0\}$ is especially important
for our purpose.  By \eqref{weyl} and \eqref{reflection} we have
apparently
\begin{equation}
  c(0,w\gamma)=c(0,\gamma)\,,\quad {}^\forall w\in W\,,
\end{equation}
and
\begin{equation}
  c(0,-\gamma)=c(0,\gamma)\,.
\end{equation}
Let us introduce
\begin{equation}\label{defsofI}
  \mathcal{I}_0 := \frac{1}{2}\sum_\gamma c(0,\gamma)\,,\qquad
  \mathcal{I}_{2n} := \sum_{\gamma>0}c(0,\gamma)(\gamma,\gamma)^n\,,
  \quad (n\ge 1)\,.
\end{equation}
We expand $E_2\phi_{k,m}$ as
\begin{equation}
(E_2\phi_{k,m})(\tau,z)=\sum\limits_{n,\gamma} \tilde
c(n,\gamma)q^n\zeta^\gamma\,,
\end{equation}
and define $\tilde P_0$, $\tilde \mathcal{I}_0$ and $\tilde
\mathcal{I}_{2n}$ $(n \in \NN)$  with the obvious
replacement of $c(0,\gamma)$ by $\tilde c(0,\gamma)$.

Since $W$ acts irreducibly on $\mathfrak{h}_\RR$, an argument similar
to [Bo1, Lemma 6.1] implies that for any $v\in \mathfrak{h}_\RR$,
$\sum_{\gamma>0}c(0,\gamma)(\gamma,v)^{2n}$ $(n=1,2,\ldots)$ is equal
to $(v,v)^n$ up to a multiplicative constant. This should also be true
for $v\in \mathfrak{h}_\CC$.  Thus the distribution of $P_0$ is
`rotationally symmetric'.  Concretely we will use
\begin{equation}\label{indices}
  \begin{split}
  \sum_{\gamma>0}c(0,\gamma)(\gamma,v)^2 &= \frac{\mathcal{I}_2}{s}\,
  (v,v)\,,\\
    \sum_{\gamma>0}c(0,\gamma)(\gamma,v)^4 &=
  \frac{3\mathcal{I}_4}{s(s+2)}\, (v,v)^2\,,  
  \end{split}
\end{equation}
for any $v\in \mathfrak{h}_\CC$. 
Similarly we have
\begin{equation}\label{indextilde}
  \sum_{\gamma>0}\tilde c(0,\gamma)(\gamma,v)^2=\frac{\tilde
  \mathcal{I}_2}{s}\, (v,v)\,,\quad {}^\forall v\in \mathfrak{h}_\CC\,.
\end{equation}

Let $z_1,\ldots,z_s$ be the components of $z\in \mathfrak{h}_\CC$ when
expressed in terms of the standard orthonormal basis.  Set
$\zeta_i=\mathbf{e}[z_i]$ $(i=1,2,\ldots,s)$. As a little calculation
shows, the differential operator
\begin{equation}
  \mathcal{D}_{k,m}:=q\frac{\partial}{\partial q}-
  \frac{1}{2m}\sum_{i=1}^s\left(
    \zeta_i\frac{\partial}{\partial\zeta_i}\right)^2
  +\frac{s-2k}{24}E_2(\tau)\,,
\end{equation}
increases the modular weight by two, {\it i.e.}
\begin{equation}
  \mathcal{D}_{k,m}(J_{k,m}^{\mathrm{nh}}(\mathfrak{g})) \subset
  J_{k+2,m}^{\mathrm{nh}}(\mathfrak{g})\,.
\end{equation}

A simple but useful observation [Bo1, Lemma 9.2] is that the constant
term in the $q$-expansion of a nearly holomorphic modular form of
weight $2$ vanishes. Hence if we let the symbol $\vert_{q^0}$ mean
picking up the constant term in the $q$-expansion, we have
$(\mathcal{D}_{0,m}\phi_{0,m})(\tau,0)\vert_{q^0}=0$, from which it
follows that
\begin{equation}\label{zero}
  \frac{1}{24}(E_2\phi_{0,m})(\tau,0)\vert_{q^0} =
  \frac{\mathcal{I}_2}{ms}\,,\quad(k=0)\,,
\end{equation}
or equivalently,
\begin{equation}
  \sum_{\substack{n>0\\ \gamma}}\sigma_1(n)c(-n,\gamma)=
\frac{\mathcal{I}_0}{12}-\frac{\mathcal{I}_2}{ms}\,, \quad(k=0)\,.
\end{equation}

Similarly 
$(E_4\phi_{-2,m})(\tau,0)\vert_{q^0}=0$ implies
\begin{equation}
  \sum_{\substack{n>0\\ \gamma}}\sigma_3(n)c(-n,\gamma) = -
  \frac{\mathcal{I}_0}{120}\,, \quad (k=-2)\,.
\end{equation}
It is also not difficult to see 
\begin{equation}
  \begin{split}\label{minus2-1}
    \frac{1}{24}({E_2}^2\phi_{-2,m})(\tau,0)\vert_{q^0}& =
    \frac{1}{2}\left(q\frac{dE_2}{dq}\phi_{-2,m}
    \right)(\tau,0)\vert_{q^0}\\
    &=-12 \sum_{\substack{n>0\\ \gamma}} n \sigma_1(n)c(-n,\gamma)\,,
    \quad (k=-2)\,,
  \end{split}
\end{equation}
where we used
\begin{equation}
  q\frac{d E_2}{dq} =\frac{{E_2}^2-E_4}{12}\,.
\end{equation}
On the other hand,
$(\mathcal{D}_{0,m}\mathcal{D}_{-2,m}\phi_{-2,m})(\tau,0)\vert_{q^0}=0$
leads after a small algebra to
\begin{equation}\label{minus2-2}
  \frac{1}{24}({E_2}^2\phi_{-2,m})(\tau,0)\vert_{q^0} = \frac{2\tilde
    \mathcal{I}_2}{ms} - \frac{12\mathcal{I}_4}{m^2s(s+2)}\,,\quad
  (k=-2)\,.
\end{equation}
Therefore combining \eqref{minus2-1} and \eqref{minus2-2} we obtain
\begin{equation}
  \sum_{\substack{n>0\\ \gamma}} n \sigma_1(n)c(-n,\gamma)= -
  \frac{\tilde \mathcal{I}_2}{6ms} +
  \frac{\mathcal{I}_4}{m^2s(s+2)}\,,\quad (k=-2) \,,
\end{equation}
or equivalently,
\begin{equation}
  \sum_{\substack{n>0\\ \gamma}}\{msn-2(\gamma,\gamma)\}
  \sigma_1(n)c(-n,\gamma) = - \frac{\mathcal{I}_2}{6} +
  \frac{\mathcal{I}_4}{m(s+2)}\,,\quad (k=-2)\,.
\end{equation}

\section{Theta correspondence, real polylogarithms and 
 the Eisenstein-{\allowbreak}Kronecker  double series}

Any nearly holomorphic Weyl-invariant Jacobi form $\phi_{k,m}\in
J_{k,m}^{\mathrm{nh}}(\mathfrak{g})$ can be expanded as
\begin{equation}\label{decoupling}
  \phi_{k,m}(\tau,z)=\sum_{\alpha\in P/mQ^\vee}h_\alpha(\tau)\,
  \theta_{\alpha,m}(\tau,z)\,,
\end{equation}
where
\begin{equation}
  \theta_{\alpha,m}(\tau,z):= \sum_{\gamma\in \alpha + mQ^\vee}
  q^{\frac{1}{2m}(\gamma,\gamma)} \zeta^\gamma\,,
\end{equation}
are the theta functions associated with the integrable representations
of the affine Lie algebra [Kac].  In this section, we fix
$\phi_{k,m}\in J_{k,m}^{\mathrm{nh}}(\mathfrak{g})$ for a simple Lie
algebra $\mathfrak{g}$ of rank $s$, together with its expansion
\eqref{decoupling}. We assume that $k$ is a non-positive even integer.

Let $L$ and $M$ be  the indefinite lattices defined by
\begin{equation}
  L:=H(-m)_2\oplus P,\qquad M:=H(-m)_1\oplus L\,,
\end{equation}
where $H(-m)_1=\ZZ e_1+\ZZ e_2$ and $H(-m)_2=\ZZ f_1+\ZZ f_2$ are
mutually isomorphic indefinite lattices of rank two with
$((e_i,e_j))=((f_i,f_j))=\bigl(\begin{smallmatrix} 0&-m\\-m&0
  \end{smallmatrix}\bigr)$.

  Then $M$ can be decomposed as
\begin{equation}
  M=\bigcup_{\alpha\in P/mQ^\vee}M_\alpha,\quad M_\alpha :=
  H(-m)_1\oplus H(-m)_2\oplus (\alpha + mQ^\vee)\,.
\end{equation}
Let $\DD$ be one of the two connected components of 
\begin{equation}
  \{[\omega]\in \PP(M_\CC) \mid \text{$(\omega,\omega)=0$ and
    $(\omega,\bar \omega)<0$}\}\,,
\end{equation}
where $[\omega]$ is a line through $\omega\in M_\CC \smallsetminus
\{0\}$.

 The cone $\{x\in L_\RR \mid
(x,x)<0\}$ consists of two connected components. We fix one of them
(the future light-cone) and denote it by $C^+(L_\RR)$.
Then the {\it tube domain\/}
\begin{equation}
  \mathcal{H}:=L_\RR+\sqrt{-1}C^+(L_\RR)\subset
  L_\CC\,,
\end{equation} gives a realization of the type IV domain $\DD$ with 
the isomorphism $\mathcal{H}\xrightarrow{\sim}\DD$ given by
\begin{equation}\label{iso1}
  y \in \mathcal{H} \longrightarrow [\omega(y)]\in \DD\,,\quad
  \mathrm{with}\quad \omega(y)=e_1+\frac{(y,y)}{2m}e_2+y\,.
\end{equation}
We will use this isomorphism freely in the following.

We also always parametrize the elements of $\mathcal{H}$ as
\begin{equation}
  \mathcal{H} \owns y=\sigma f_1+\tau f_2+mz\,, \quad z\in K_\CC\,.
\end{equation}
Thus if we set 
\begin{equation}
  Y:= \Ima\sigma\Ima \tau-\frac{m}{2}(\Ima z,\Ima z)\,,
\end{equation}
we find that
\begin{equation}
  (\omega,\bar \omega)=2(\Ima y,\Ima y )=-4mY<0\,.
\end{equation}

Each element $[\omega]\in \DD$ determines a negative definite two
dimensional vector subspace of $M_\RR$ with an oriented basis.  Given
a lattice element $\lambda\in M$ we denote by $\lambda_-^{[\omega]}$
the projection of $\lambda$ onto this vector subspace and set
$\lambda_+^{[\omega]}:=\lambda - \lambda_-^{[\omega]}$. Therefore we
have a relation:
\begin{equation}
  \frac{1}{2}\bigl(\lambda_-^{[\omega]}, \lambda_-^{[\omega]}\bigr) =
  \frac{\abs{(\lambda,\omega)}^2}{(\omega,\bar \omega)}\,.
\end{equation}

Let $\mathbb{F}$ be the fundamental domain of the modular group
$SL_2(\ZZ)$ in $\HH$. We put $v_1:=\Rea{\rho}$, $v_2:=\Ima{\rho}$ and
$q_\rho:={\bf e}[\rho]$ for $\rho\in \HH$..  Now we consider the
following theta correspondence:
\begin{equation}\label{modularintegral}
  {\cal V} = \frac{1}{4}\int_{\mathbb{F}} \frac{dv_1\,dv_2}{v_2}
  \sum_{\alpha\in P/mQ^\vee} \left( E_2(\rho) - \frac{3}{\pi v_2}
  \right)^{-k/2} h_\alpha(\rho) \Theta_{\alpha,m}(\rho, \bar \rho,
  [\omega])\,,
\end{equation}
where the indefinite theta functions $\Theta_{\alpha,m}$ are defined by
\begin{equation}
  \Theta_{\alpha,m}(\rho,\bar \rho, [\omega]):= \sum_{\lambda\in
    M_\alpha} q_\rho^{\frac{1}{2m} \bigl(\lambda_+^{[\omega]},
    \lambda_+^{[\omega]}\bigr)}\bar q_\rho^{-
    \frac{1}{2m}\bigl(\lambda_-^{[\omega]},
    \lambda_-^{[\omega]}\bigr)}\,.
\end{equation}
This is a generalization of [HaM] and a special case of [Bo2]. The
modular integral \eqref{modularintegral} is divergent and it must be
regularized appropriately. The evaluation of \eqref{modularintegral}
is now a standard task, however here we wish to emphasize that the final
result can be neatly expressed in terms of the real analytic
polylogarithms introduced in the work of Zagier [Z]. Besides
aesthetic, this gives a motivation for the construction to be explained in
the next section.

Thus we first collect relevant materials from [Z] with a slight change
of the notation.  For a positive odd integer $r$ we define
\begin{equation}
  \Lir_r(\xi):=(-1)^{a-1}\sum_{j=a}^r 2^{1-j}\binom{j-1}{a-1}
  \frac{(-\log\abs{\xi})^{r-j}}{(r-j)!}\Rea(\Li_j(\xi))\,,
\end{equation}
where $r=2a-1$. This differs from $D_{a,a}$ in [Z] only by a
multiplicative constant. $\Lir_r$ is a single-valued real analytic
function on $\CC\smallsetminus [1,\infty)$ satisfying the inversion
relation
\begin{equation}\label{inversionreal}
  \Lir_r(\xi^{-1})=\Lir_r(\xi)+\frac{(\log\abs{\xi})^r}{r!}\,.
\end{equation}
 For instance, we have
\begin{align}
  &\Lir_1(\xi)=\Rea(\Li_1(\xi))\,,\\ &\Lir_3(\xi)= \frac{1}{2}
  \Rea\bigl[ \log \abs{\xi} \cdot \Li_2(\xi) - \Li_3(\xi) \bigr]\,.
\end{align}

The Bernoulli polynomials $B_n(x)$ ($n\in \NN_0)$ are monic
polynomials of degree $n$ defined through
\begin{equation}
  \frac{te^{xt}}{e^t-1}=\sum_{n=0}^\infty B_n(x)\frac{t^n}{n!}\,.
\end{equation}
We need the first five of them and they  are given explicitly by
\begin{align}
  &B_0(x)=1\,, \nonumber\\ 
  &B_1(x)=x-\frac{1}{2}\,, \nonumber\\
 &B_2(x)=x^2-x+\frac{1}{6}\,, \\
 &B_3(x)=x^3-\frac{3}{2}x^2+\frac{1}{2}x\,, \nonumber\\
 & B_4(x)=x^4-2x^3+x^2-\frac{1}{30}\,. \nonumber
\end{align}
The Bernoulli polynomials are related to the Bernoulli numbers by
\begin{equation}
  B_n(x)=\sum_{k=0}^n \binom{n}{k}B_kx^{n-k}\,,
\end{equation}
so that
\begin{equation}\label{bernoulli}
  B_{2n}(x)-B_{2n}(-x)=-2nx^{2n-1}\,.
\end{equation}

Zagier [Z] introduced the following elliptic extension of
$\Lir_r(\xi)$:
\begin{equation}
  \Lir_r(q,\xi):=\sum_{n=0}^\infty \Lir_r(q^n\xi)+ \sum_{n=1}^\infty
  \Lir_r(q^n\xi^{-1})-\mathbb{X}_r(q,\xi)\,,
\end{equation}
where
\begin{equation}\label{Xr}
  \begin{split}
    \mathbb{X}_r(q,\xi):=&\frac{(\log\abs{q})^r}{(r+1)!}\, B_{r+1}\!
    \left(\frac{\log\abs{\xi}}{\log\abs{q}}\right)\\ 
    =&\sum_{j=-1}^r\frac{B_{j+1}}{(r-j)!(j+1)!}
      (\log\abs{\xi})^{r-j}(\log\abs{q})^j\,.
    \end{split}
\end{equation}
Note that $\mathbb{X}_r(q,\xi)$ would be a polynomial in
$\log\abs{\xi}$ and $\log\abs{q}$ if the summation started from $j=0$.

To evaluate $\mathcal{V}$ we have to divide it into three parts [DKL]
\begin{equation}
  \mathcal{V}=\mathcal{V}_0+\mathcal{V}_1+\mathcal{V}_2\,,
\end{equation}
where $\mathcal{V}_0$, $\mathcal{V}_1$ and $\mathcal{V}_2$ correspond
respectively to the zero orbit, the degenerate orbits and the
non-degenerate orbits of the modular group.  The Eisenstein-Kronecker
double series $E_t(q,\xi)$ $(\Rea t >1)$ is defined as
\begin{equation}
  E_t(q,\xi):=\sum_{\substack{w\in \ZZ+\tau\ZZ\\ w\ne 0}}
\frac{\chi_w(\tau,x)}{\abs{w}^{2t}}\,,
\end{equation}
where $q=\mathbf{e}[\tau]$, $\xi=\mathbf{e}[x]$ and
\begin{equation}
  \chi_w(\tau,x):={\bf e} \left[ \frac{\Ima(\bar w x)}{\Ima\tau}
  \right]\,.
\end{equation}
The Eisenstein-Kronecker double series appear quite naturally in the
evaluation of $\mathcal{V}_1$. See for instance [Kaw2].  Interestingly
Zagier found that $\Lir_r(q,\xi)$ coincides modulo some factor with
$E_a(q,\xi)$. The case $r=1=a$ reduces to the classical Kronecker
limit formula and as usual we have to make an analytic continuation
and discard the divergent part. Consequently $\mathcal{V}_1$ can be
expressed in terms of the real elliptic polylogarithms $\Lir_r(q,\xi)$
where $1 \le r \le 1-k$.

A more detailed explanation is as follows. Suppose $(\ell,n,\gamma)\in
\ZZ\times\ZZ\times P$. Then, by $(\ell,n,\gamma)>0$ we mean either of
the three cases:
$$ \mathrm{(i)}\ \ell>0\,, \quad \mathrm{(ii)}\ \ell=0\,,\  n>0\,, \quad
\mathrm{(iii)}\ \ell=n=0\,,\  \gamma>0\,.$$ 
Using the relations between $E_a(q,\xi)$ and $\Lir_r(q,\xi)$,
$\mathcal{V}_1$ is a linear combination of $\inv{2} \sum_\gamma
c_r(0,\gamma) \Lir_r(q,\zeta^\gamma)$ $(1\le r \le 1-k)$ where the
coefficients $c_r(0,\gamma)$ are determined through the expansions
\begin{equation}
  ({E_2}^{\frac{1-k-r}{2}}\phi_{k,m})(\tau,z)
  =\sum_{n,\gamma}c_r(n,\gamma)q^n\zeta^\gamma\,.
\end{equation}
By applying  \eqref{inversionreal} and \eqref{bernoulli} it is
straightforward to show
\begin{equation}\label{degenerate}
  \begin{split}
  \inv{2} \sum_\gamma c_r(0,\gamma) \Lir_r(q,\zeta^\gamma)&=
   \frac{c_r(0,0)}{2}\Lir_r(1)+
    \sum_{(0,n,\gamma)>0}c_r(0,\gamma)\Lir_r(q^n\zeta^\gamma)\\ 
    &-\frac{c_r(0,0)}{2}\frac{B_{r+1}}{(r+1)!}(\log\abs{q})^r-
    \sum_{\gamma>0}c_r(0,\gamma)\mathbb{X}_r(q,\zeta^\gamma)\,.
\end{split}
\end{equation}
As mentioned we have to be careful about the case $r=1$ since
$\Lir_1(1)$ is divergent and a suitable regularization is needed. We
will choose a regularization scheme such that $\Lir_1(1)$ is formally
replaced by $-\inv{2}\log \hat Y$ where
\begin{equation}
  \begin{split}
    \hat Y:=&\log\abs{p}\log\abs{q} -
    \frac{m}{2}(\log\abs{\zeta},\log\abs{\zeta})\\ 
    =&(2\pi)^2 Y\,,
  \end{split}
\end{equation}
with $\log\abs{\zeta}:=-2\pi \Ima z$.

If we include the contribution from $\mathcal{V}_2$,
$\sum_{(0,n,\gamma)>0} c_r(0,\gamma)\Lir_r(q^n\zeta^\gamma)$ in
\eqref{degenerate} is replaced by $\sum_{(\ell,n,\gamma)>0}c_r(\ell
n,\gamma)\Lir_r(p^\ell q^n\zeta^\gamma)$. Therefore the final result
of the evaluation of $\mathcal{V}$ can be written in terms of $\Lir_r$
$(1\le r \le 1-k)$.

In the following we summarize the results of the calculations for
$k=0$ and $k=-2$ since these cases are of direct interest in later
sections. However, in principle one may work out for other cases. To
reach the results we have frequently used the various identities given
in \S 2. In particular the terms in $\mathcal{V}_1$ stemming from the
summand with $j=-1$ in \eqref{Xr} are cancelled against
$\mathcal{V}_0$.  (In [Bo2] it was shown that such cancellations must
always occur by quite a general argument.)

\medskip
\underline{$k=0$}

\begin{equation}
  \mathcal{V}_0 = -\frac{\hat Y}{24\log\abs{q}}
  (E_2\phi_{0,m})(\tau,0)\vert_{q^0}
\end{equation}

\begin{equation}
  \mathcal{V}_1=\inv{2} \sum_\gamma c(0,\gamma) \Lir_1(q,\zeta^\gamma)
\end{equation}

\begin{equation}
  \begin{split}
    \mathcal{V}=-\frac{\mathcal{I}_2}{ms}\log
    \abs{p}-\frac{\mathcal{I}_0}{12}\log \abs{q} +
    \inv{2}\sum_{\gamma>0} c(0,\gamma)(\gamma,\log \abs{\zeta})\\ -
    \frac{c(0,0)}{4}\log \hat Y +\sum_{(\ell,n,\gamma)>0}c(\ell
    n,\gamma)\Lir_1(p^\ell q^n\zeta^\gamma)
  \end{split}
\end{equation}

\underline{$k=-2$}

\begin{equation}
  \mathcal{V}_0= -\frac{\hat Y}{48\log\abs{q}}
  ({E_2}^2\phi_{-2,m})(\tau,0) \vert_{q^0}
\end{equation}

\begin{equation}
  \mathcal{V}_1=\inv{2} \sum_\gamma \tilde c(0,\gamma)
  \Lir_1(q,\zeta^\gamma) + \frac{12}{\hat Y}\cdot \inv{2}\sum_\gamma
  c(0,\gamma) \Lir_3(q,\zeta^\gamma)
\end{equation}

\begin{equation}\label{modularminustwo}
  \begin{split}
    \mathcal{V}=\left(-\frac{\tilde
        \mathcal{I}_2}{ms}+\frac{6\mathcal{I}_4}{m^2s(s+2)}\right)\log
    \abs{p} -\frac{\tilde \mathcal{I}_0}{12}\log
    \abs{q}+\inv{2}\sum_{\gamma>0}\tilde
    c(0,\gamma)(\gamma,\log\abs{\zeta})\\ -\frac{\tilde c(0,0)}{4}\log
    \hat Y +\sum_{(\ell,n,\gamma)>0}\tilde c(\ell
    n,\gamma)\Lir_1(p^\ell q^n \zeta^\gamma)\\ +\frac{12}{\hat
      Y}\Bigl[-\frac{\mathcal{I}_4}{4ms(s+2)} \log \abs{p}\cdot(\log
    \abs{\zeta},\log\abs{\zeta})+ \frac{\mathcal{I}_0}{720}(\log
    \abs{q})^3\\ -\frac{\mathcal{I}_2}{24s}\log
    \abs{q}\cdot(\log\abs{\zeta},\log\abs{\zeta})
    +\frac{1}{12}\sum_{\gamma>0}c(0,\gamma)(\gamma,\log
    \abs{\zeta})^3\\ -\frac{c(0,0)}{4}\zeta(3) +
    \sum_{(\ell,n,\gamma)>0} c(\ell n,\gamma) \Lir_3(p^\ell
    q^n\zeta^\gamma)\Bigr]\\ 
  \end{split}
\end{equation}

\section{The Borcherds lifting and  the 
Beilinson-Levin elliptic polylogarithms}

The main purpose of this section is to introduce a lifting procedure
for a nearly holomorphic Weyl-invariant Jacobi form whose weight is a
non-positive even integer. This is a generalization of the lifting
which was introduced by Borcherds [Bo1] for the case of weight $0$ and
was applied for the construction of automorphic forms on type IV
domains as infinite products [Bo1,GN1,GN2]. (To avoid a confusion,
we should remark in advance that we will not take the ``exponential''
unlike [Bo1,GN1,GN2].) As mentioned in Introduction, our lifting
procedure involves the elliptic polylogarithms of Beilinson and Levin
[BeL]. Thus we begin by reviewing these functions. For a more precise
formulation we refer to [BeL] and especially to \S 4.4 of that paper.

The polylogarithms defined by \eqref{polylog} 
satisfy 
\begin{equation}\label{polylogrel}
  \xi \frac{\partial}{\partial \xi}\Li_r(\xi)=\Li_{r-1}(\xi)\,,\quad
  r\in \NN\,,
\end{equation}
with $\Li_0(\xi)=\xi/(1-\xi)$. In other words we have
\begin{equation}
  \Li_r(\xi)=\int^\xi_0 \Li_{r-1}(\xi')\frac{d\xi'}{\xi'}\,,\quad r\in \NN\,.
\end{equation}
Using this relation inductively, each polylogarithm can be
analytically continued to a multi-valued holomorphic function on
$\PP^1\smallsetminus\{0,1,\infty \}=\CC\smallsetminus \{0,1\}$. 

The monodromy of the polylogarithms has been studied in [R]. See also
[H] for instance. Consider the $(r+1)^\mathrm{th}$ order differential
equation
\begin{equation}
  \frac{d}{d\xi}\left[\frac{1-\xi}{\xi} \left(\xi\frac{d}{d\xi}\right)^r
    f(\xi)\right]=0\,,
\end{equation}
whose ``local'' solution space is spanned by $\Li_r$,
$S_{r-1},S_{r-2},\ldots,S_0$. Here we have set
$S_{r-j}=\frac{(2\pi\sqrt{-1})^j}{(r-j)!}(\log \xi )^{r-j}$, $(1\le j
\le r)$.  Under the monodromy transformations corresponding to the
elements of $\pi_1(\CC\smallsetminus\{0,1\},*)$ where $*$ is a base
point, $\Li_r(\xi)$ generally shifts by $\QQ$-linear combinations of
$S_{r-1},S_{r-2},\ldots,S_0$. Thus, in order to kill off the monodromy
and attain the effective single-valuedness we are led to define
$\Lic_r(\xi)$ as $\Li_r(\xi)$ modulo any $\QQ$-linear combinations of
$S_{r-1},S_{r-2},\ldots,S_0$. In the following we frequently write
down equations involving $\Lic_r$ with various different arguments. It
should be understood that these are valid only modulo $\QQ$-linear
combinations of $S_{r-1},S_{r-2},\ldots,S_0$ where
$S_{r-1},S_{r-2},\ldots,S_0$ are determined according to the argument
of $\Lic_r$.

 For instance we have $\Lic_r(1)=\zeta(r)$ $(r>1)$ since
$\Li_r(1)$ $(r>1)$ coincides with $\zeta(r)$ modulo
$\frac{(2\pi\sqrt{-1})^r}{(r-1)!} \ZZ$. (See [H].)  Actually
$\Lic_r(1)=0$ if $r$ is even, since $\zeta(r)=-\frac{(2\pi
  \sqrt{-1})^{r}}{2 \cdot r!}B_r \in (2\pi \sqrt{-1})^{r} \QQ$ in that
case.

By an induction argument using \eqref{polylogrel} one can show that
the ordinary polylogarithms satisfy the inversion formula
\begin{equation}\label{inversionordinary}
  (-1)^{r-1}\Li_r(\xi^{-1}) = \Li_r(\xi) +
  \sum_{j=0}^r\frac{B_j (2\pi\sqrt{-1})^j}{(r-j)!j!} (\log\xi)^{r-j}\,,
\end{equation}
where we have chosen the branch of $\Li_r$ so that $\lim_{\xi
  \rightarrow 1}\Li_r(\xi^{\pm 1})=\zeta(r)$.
Then it is easy to see that 
\begin{equation}\label{inversioncomplex}
  (-1)^{r-1}\Lic_r(\xi^{-1}) = 
\Lic_r(\xi)+\frac{(\log\xi)^r}{r!}\,.
\end{equation}
More directly one can show \eqref{inversioncomplex} by induction since
\eqref{polylogrel} is true if we replace $\Li$ by $\Lic$.  

Notice that if $r$ is odd, \eqref{inversioncomplex} is the same type
of transformation law as \eqref{inversionreal} for the real analytic
single-valued polylogarithm $\Lir_r$.  Henceforth we almost always
assume that $r$ is a positive odd integer to render the formulas in
this section in parallel with those in the previous section.

In the course of their motivic interpretation of Zagier's
results [Z], Beilinson and Levin [BeL] introduced the elliptic
polylogarithm:
\begin{equation}\label{BLellpoly}
  \Lic_r(q,\xi):=\sum_{n=0}^\infty \Lic_r(q^n\xi)+ \sum_{n=1}^\infty
  \Lic_r(q^n\xi^{-1})-\mathcal{X}_r(q,\xi)\,,
\end{equation}
where
\begin{equation}
  \begin{split}
    \mathcal{X}_r(q,\xi):=&\frac{(\log{q})^r}{(r+1)!}\, \tilde
    B_{r+1}\!  \left(\frac{\log{\xi}}{\log{q}}\right)\\ 
    =&\sum_{j=0}^r\frac{B_{j+1}}{(r-j)!(j+1)!}
      (\log{\xi})^{r-j}(\log{q})^j\,,
    \end{split}
\end{equation}
with
\begin{equation}
  \tilde B_n(x):=B_n(x)-x^n\,.
\end{equation}
The RHS of \eqref{BLellpoly} can be obtained from the formal sum
$\sum_{n\in \ZZ}\Lic_r(q^n\xi)$ by using \eqref{inversioncomplex} and
then applying the $\zeta$ function regularization.
Obviously $\Lic_r(q,\xi)$ is an analog of $\Lir_r(q,\xi)$.  However,
notice that this time $\mathcal{X}_r(q,\xi)$ is a polynomial of
$\log{\xi}$ and $\log{q}$.

The key devices in the lifting procedure are the Hecke operators.
Given a $\phi_{k,m}(\tau,z)=\sum_{n,\gamma} c(n,\gamma)
q^n\zeta^\gamma\in J_{k,m}^{\rm nh}(\mathfrak{g})$, the actions of the
Hecke operators $V_\ell$ $(\ell \in \NN)$ are defined by [EZ]
\begin{equation}
  \left( \phi_{k,m}\vert V_\ell\right) (\tau,z) :=
  \ell^{k-1}\sum_{ad=\ell} \sum_{b=0}^{d-1} d^{-k}
  \phi_{k,m}\left(\frac{a\tau+b}{d},az\right) \in J_{k,\ell m}^{\rm
    nh}(\mathfrak{g}),
\end{equation}
where the first sum runs over all the  positive integers
$a, d$ such that $ad=\ell$.

We can formally extend the definition of the polylogarithm
$\Li_r(\xi)$ for all $r\in \ZZ$ by repeatedly using
\eqref{polylogrel}.  Then we see that $\Li_r(\xi)$ $(r\le 0)$ are
rational functions with poles at $\xi=1$ and satisfy
\begin{equation}\label{negativeinv}
  (-1)^{r-1}\Li_r(\xi^{-1})=\Li_r(\xi)\,,\quad r \le 0\,.
\end{equation}
A crucial identity for us is
\begin{equation}\label{nondeghecke}
  \sum_{\ell=1}^\infty p^\ell \left( \phi_{k,m}\vert V_\ell\right)
  (\tau,z) = \sum_{\substack{\ell,n,\gamma\\ \ell>0}} c(\ell n,\gamma)
  \Li_{1-k}(p^\ell q^n\zeta^\gamma)\,,
\end{equation}
which can be shown by a little calculation and is valid for all $k\in
\ZZ$. Thus the lifting procedure is directly related to
polylogarithms.  Eq.\eqref{nondeghecke} with $k=1-r$ roughly
corresponds to the non-degenerate part $\mathcal{V}_2$ in \S 3.

The  part which is independent of $p$ (corresponding to
$\mathcal{V}_1$) may be constructed by considering, in analogy to
\eqref{degenerate}, the weighted sum of elliptic
polylogarithms  over $P_0$:
\begin{equation}
  \begin{split}
    \inv{2}\sum_\gamma c(0,\gamma)
    \Lic_r(q,\zeta^\gamma)&=\frac{c(0,0)}{2}\zeta(r) +
   \sum_{(0,n,\gamma)>0}c(0,\gamma)\Lic_r(q^n\zeta^\gamma)\\ 
    &-\frac{c(0,0)}{2}\frac{B_{r+1}}{(r+1)!}(\log {q})^r-
    \sum_{\gamma>0}c(0,\gamma)\mathcal{X}_r(q,\zeta^\gamma)\,,
\end{split}
\end{equation}
where $k=1-r$ and we used again the inversion relation
\eqref{inversioncomplex}. 
This prompts  us to  define the Hecke operator $V_0$ by
\begin{equation}
  \left( \phi_{k,m}\vert V_0\right) (\tau,z) :=
  \frac{c(0,0)}{2}\zeta^*(1-k) +
  \sum_{(0,n,\gamma)>0}c(0,\gamma)\Li_{1-k}(q^n\zeta^\gamma)\,,
\end{equation}
where $\zeta^*(1)=0$ and $\zeta^*(n)=\zeta(n)$ if $n$ is a nonzero
integer.

Thus we are led to define
\begin{equation}
  \begin{split}
  {\tilde {\Lic}}_r^\phi(p,q,\zeta)&:=\sum_{\ell=0}^\infty p^\ell
  \left( \phi_{1-r,m}\vert V_\ell\right)(\tau,z)^!\\
&-\frac{c(0,0)}{2}\frac{B_{r+1}}{(r+1)!}(\log {q})^r-
    \sum_{\gamma>0}c(0,\gamma)\mathcal{X}_r(q,\zeta^\gamma)\,,
  \end{split}
\end{equation}
where the symbol $!$ means that we should replace $\Li_r$ by $\Lic_r$
when applying \eqref{nondeghecke}.

We wish to define $\Lic_r^\phi(p,q,\zeta)$ by adding to ${\tilde
  {\Lic}}_r^\phi(p,q,\zeta)$ the part which will consist of terms depending
on $\log p$.  For $r=1$ and $r=3$ the following definitions seem to be
working:
\begin{equation}
  \Lic_1^\phi(p,q,\zeta):=-\frac{\mathcal{I}_2}{ms}\log p +{\tilde
    {\Lic}}_1^\phi(p,q,\zeta)\,,
\end{equation}
and
\begin{equation}
  \begin{split}
    \Lic_3^\phi(p,q,\zeta) :=- \frac{\mathcal{I}_4}{4ms(s+2)} \log p
    \cdot(\log \zeta,\log\zeta)+{\tilde {\Lic}}_3^\phi(p,q,\zeta)\,.
  \end{split}
\end{equation}
More explicit expressions are given by
\begin{equation}\label{formLic_1}
  \begin{split}
    \Lic_1^\phi(p,q,\zeta)=-\frac{\mathcal{I}_2}{ms}\log
    p-\frac{\mathcal{I}_0}{12}\log q + \inv{2}\sum_{\gamma>0}
    c(0,\gamma)(\gamma,\log \zeta)\\ 
    +\sum_{(\ell,n,\gamma)>0}c(\ell
    n,\gamma)\Lic_1(p^\ell q^n\zeta^\gamma)\,,
  \end{split}
\end{equation}
and
\begin{equation}\label{formLic_3}
  \begin{split}
    \Lic_3^\phi(p,q,\zeta) =-\frac{\mathcal{I}_4}{4ms(s+2)} \log
    p\cdot(\log \zeta,\log\zeta)+ \frac{\mathcal{I}_0}{720}(\log
    q)^3\\
-\frac{\mathcal{I}_2}{24s}\log q\cdot(\log\zeta,\log\zeta)  
    +\frac{1}{12}\sum_{\gamma>0}c(0,\gamma)(\gamma,\log \zeta)^3\\
    +\frac{c(0,0)}{2}\zeta(3) + \sum_{(\ell,n,\gamma)>0} c(\ell
    n,\gamma) \Lic_3(p^\ell q^n\zeta^\gamma)\,,
  \end{split}
\end{equation}
where $\log\zeta:=2\pi \sqrt{-1} z$. It should be kept in mind that
the $c(n,\gamma)$'s are the coefficients of $\phi_{0,m}$ in
\eqref{formLic_1} while they are the coefficients of $\phi_{-2,m}$ in
  \eqref{formLic_3}.

The choices of $\log p$ dependent terms are apparently motivated by
the calculations in the previous section and may look ad hoc within
the logic of this section. However we can show that with these choices
$\Lic_1^\phi(p,q,\zeta)$ and $\Lic_3^\phi(p,q,\zeta)$ have nice wall-crossing
behaviors under the exchange of $p$ and $q$ which determine part of
the (quasi-)automorphic properties of $\Lic_1^\phi(p,q,\zeta)$ and
$\Lic_3^\phi(p,q,\zeta)$.  In order to know these behaviors we must use
various identities derived in \S 2.  For $\Lic_1^\phi(p,q,\zeta)$ we have
\begin{equation}\label{cross1}
  \Lic_1^\phi(p,q,\zeta)=\Lic_1^\phi(q,p,\zeta)\,,
\end{equation}
by using
\begin{equation}
  \begin{split}
    -\sum_{\substack{\ell>0\\ n<0\\ \gamma}} c(\ell n,\gamma)
    \log(p^\ell q^n\zeta^\gamma)&= - \sum_{\substack{n>0\\ \gamma}}
    \sigma_1(n) c(-n,\gamma) (\log p-\log q)\\ 
    &=\left(-\frac{\mathcal{I}_0}{12} +
      \frac{\mathcal{I}_2}{ms}\right)(\log p-\log q)\,.
 \end{split}
 \end{equation}
 On the other hand we have
\begin{equation}\label{cross3}
  \begin{split}
    &\Lic_3^\phi(p,q,\zeta)-\Lic_3^\phi(q,p,\zeta)\\ 
    &=\left( - \frac{\tilde \mathcal{I}_2}{12ms} +
      \frac{\mathcal{I}_4}{2m^2s(s+2)} \right) (\log p-\log
    q)\left\{\log p \log q-\frac{m}{2}(\log\zeta,\log
      \zeta)\right\}\,.
\end{split}
\end{equation}
This can be proved by using
\begin{multline}
    -\sum_{\substack{\ell>0\\ n<0\\ \gamma}} c(\ell n,\gamma)
    \frac{(\log(p^\ell q^n\zeta^\gamma))^3}{3!} 
    = -\frac{1}{6} \sum_{\substack{n>0\\ \gamma}} \sigma_3(n)
    c(-n,\gamma) ((\log p)^3-(\log q)^3)\\  -\frac{1}{2s}
    \sum_{\substack{n>0\\ \gamma}} \sigma_1(n) (\gamma,\gamma)
    c(-n,\gamma) (\log p-\log q) (\log\zeta,\log \zeta)\\ 
     +\inv{2}\sum_{\substack{n>0\\ \gamma}} n \sigma_1(n)
    c(-n,\gamma) (\log p-\log q) \log p \log q\,, \\ 
  \end{multline}
and noting that the RHS can be rewritten as
\begin{equation}
  \begin{split}
   & \frac{\mathcal{I}_0}{720} ((\log p)^3-(\log q)^3))\\
    & + \left(-
      \frac{\mathcal{I}_2}{24s}+\frac{\mathcal{I}_4}{4ms(s+2)}\right)
    ( \log p-\log q) (\log\zeta,\log \zeta)\\ 
    & + \left( -\frac{\tilde \mathcal{I}_2}{12ms} +
      \frac{\mathcal{I}_4}{2m^2s(s+2)} \right) (\log p-\log q)
    \left\{\log p \log q-\frac{m}{2}(\log\zeta,\log \zeta) \right\}\,.
  \end{split}
\end{equation}

The functions $\Lic_1^\phi(p,q,\zeta)$ and $\Lic_3^\phi(p,q,\zeta)$ introduced
in the above and their properties \eqref{cross1} and \eqref{cross3}
 will play important roles in the next section.

\section{String duality and 
  enumeration of curves}

Now we can turn to the problem mentioned in Introduction.  We are
concerned with Calabi-Yau threefolds admitting fibrations.  The
existence of an algebraic fibration structure on $X$ is related to
that of a nef effective divisor $D$ on $X$ and the allowed types of
fibrations are classified [O] by the numerical dimension
$\nu(X,D):=\max\{n\in \NN \mid D^n \not \equiv 0\}$ where $\equiv$
stands for the numerical equivalence relation and the value of
$c_2(X)\cdot D$.  (For a nef divisor $D$ we have $c_2(X)\cdot D\ge 0$
[Mi].)

We first assume that $\nu(X,D_1)=1$ and $\rho_1>0$ which means that
$X$ allows a type $\mathrm{I}_+$ fibration [O].  In other words, there
exists a fibration $\pi_1:X\rightarrow W_1$ with $W_1\cong \PP^1$ such
that a generic non-singular fiber is an algebraic $K3$ surface.  We
have $\kappa_{11i}=0$ $({}^\forall i)$ by $\nu(X,D_1)=1$ and actually
$\rho_1=24$.  The divisor $D_1$ may be identified with a generic fiber
$K3$ of this fibration.

Another assumption we make upon $X$ is that $\nu(X,D_2)=2$ and
$\rho_2>0$. This is the case of type $\mathrm{II}_+$ fibration and
there exists an elliptic fibration $\pi_2:X\rightarrow W_2$. We assume
that this fibration has a section $\sigma_2:W_2\rightarrow X$.  The
condition $\nu(X,D_2)=2$ can be rephrased as: $\kappa_{222}=0$ but
$\kappa_{22i}\ne 0$ for some $i$.

Due to the above assumptions, $W_2$ itself is a $\PP^1$ fibration over
$\PP^1$, $\pi_{21}: W_2\rightarrow W_1$. We further assume that this
fibration is smooth. In other words, $W_2$ is a ruled surface over
$\PP^1$, hence is one of the Hirzebruch surfaces, say $\Sigma_a$
$(0\le a \le 12)$. There are two basic divisors of $\Sigma_a$, namely
$C_0$ (a section of $\pi_{21}$) and $f$ (the fiber of $\pi_{21}$) with
$C_0\cdot C_0=-a$, $f\cdot f=0$ and $C_0\cdot f=1$. There is another
section $C_\infty:=C_0+af$ satisfying $C_\infty\cdot C_\infty=a$,
$C_0\cdot C_\infty=0$ and $C_\infty \cdot f=1$.

The divisor $D_1$ may be identified with the restriction of
$\pi_2:X\rightarrow W_2$ on $f$. Thus $D_1$ is an elliptic $K3$.  On
the other hand, the divisor $D_2$ may be identified with the
restriction of $\pi_2:X\rightarrow W_2$ on $C_\infty$.  To put
differently, $q_1$ and $q_2$ are the parameters counting how many
times the images of the instanton holomorphic maps wind respectively
around $C_0$ and $f$.  If we represent $X$ in the Weierstrass form the
discriminant divisor $\Delta\subset W_2$ is given by (see for instance
[A])
\begin{equation}
  \Delta=24C_0+(24+12a)f\,.
\end{equation}
Hence $\Delta\cdot f=24$ and $\Delta\cdot C_\infty=24+12a$. This gives
immediately that $\chi(D_1)=24$ and $\chi(D_2)=24+12a$ assuming that
the singular fibers are of Kodaira type $\mathrm{I}_1$. On the other
hand, $\rho_1=c_2(D_1)$ and $\rho_2=c_2(D_2)$ in view of ${D_1}^3= 0$
and ${D_2}^3 = 0$. Therefore we have
\begin{equation}
  \rho_1=24\,,\qquad \rho_2=24+12a\,.
\end{equation}
As the divisor $D_3$ we may choose  $\sigma_2(W_2)$. 
It turns out\footnote{See the footnote 6 of [BJPS].} that 
\begin{equation}
  \rho_3=92\,.
\end{equation}

We make one more important assumption. We assume that the singular
fibers of $\pi_1:X\rightarrow W_1$ are irreducible. This ensures that
there are no extra contributions to $H_4(X,\ZZ)$ or equivalently
$H^2(X,\ZZ)$ from the singular fibers [A].

Our goal in this section is to study the behaviors of $F_0$ and $F_1$
in the limit $q_1 \rightarrow 0$ using the results obtained in the
previous sections. The main reason why we are interested in this limit
lies in its possible connection to perturbative heterotic string
compactified on $K3\times T^2$ [KLM,AL,A]. However, even if we fail to
find such a connection, the asymptotic behaviors we propose may well
be true.

Before presenting our proposal, we introduce several auxiliary
functions.  Suppose we have a simple Lie algebra $\mathfrak{g}$ of
rank $s$ and $\phi_{-2,m}(\tau,z)\in
J_{-2,m}^{\mathrm{nh}}(\mathfrak{g})$ together with the expansion
$\phi_{-2,m}(\tau,z)=\sum_{n,\gamma} c(n,\gamma)q^n\zeta^\gamma$.
Then we define
\begin{equation}
  \phi_{0,m}:=\frac{24}{s+4} \mathcal{D}_{-2,m}\phi_{-2,m}\in 
  J_{0,m}^{\mathrm{nh}}(\mathfrak{g})\,.
\end{equation}
Using these $\phi_{0,m}$ and $\phi_{-2,m}$ one can construct
$\Lic_1^\phi(p,q,\zeta)$ and $\Lic_3^\phi(p,q,\zeta)$ as explained in the
previous section. Then we can introduce
\begin{equation}
  \mathcal{F}_0(p,q,\zeta):=\Lic_3^\phi(p,q,\zeta)\,,
\end{equation}
and 
\begin{equation}
  \begin{split}
    \mathcal{F}_{0,1}(p,q,\zeta):=&\Lic_1^\phi(p,q,\zeta)\\
 & -\frac{24}{s+4}\left\{p\frac{\partial}{\partial
        p}q\frac{\partial}{\partial q} -\frac{1}{2m}\sum_{i=1}^s
      \left(\zeta_i\frac{\partial}{\partial
          \zeta_i}\right)^2\right\}\mathcal{F}_0(p,q,\zeta)\,.
\end{split}
\end{equation}
The explicit expression of $\mathcal{F}_{0,1}$ is given by
\begin{equation}
  \begin{split}
    \mathcal{F}_{0,1}= \left(-\frac{\tilde
        \mathcal{I}_2}{ms}+\frac{6\mathcal{I}_4}{m^2s(s+2)}\right)\log
    p -\frac{\tilde \mathcal{I}_0}{12}\log
    q+\inv{2}\sum_{\gamma>0}\tilde c(0,\gamma)(\gamma,\log\zeta)\\ 
    +\sum_{(\ell,n,\gamma)>0}\tilde c(\ell n,\gamma)\Lic_1(p^\ell q^n
    \zeta^\gamma)\,,
  \end{split}
\end{equation}
which should be compared  with \eqref{modularminustwo}.

The main conjecture of this paper is that by imposing an extra
condition on $X$ to be mentioned below and by selecting a suitable Lie
algebra $\mathfrak{g}$ of rank $s$ and $\phi_{-2,m}\in
J_{-2,m}^{\mathrm{nh}}(\mathfrak{g})$ with $l=h^{11}=s+3$, the
Gromov-Witten potentials $F_0$ and $F_1$ have the following asymptotic
behaviors:
\begin{align}\label{ansatz}
  F_0(q_*)&=\log u\left\{\log p\log q-\frac{m}{2}(\log
    \zeta,\log\zeta)\right\} +\mathcal{F}_0(p,q,\zeta)+O(q_1)\,,\\ 
  F_1(q_*)&=-\frac{1}{2}\cdot 24 \cdot \log u +
  \mathcal{F}_{0,1}(p,q,\zeta)+O(q_1)\,.
\end{align}
Here by setting
\begin{equation}
  t_i=\log q_i\,,\qquad (i=1,\ldots,l)\,,
\end{equation}
we should have
\begin{equation}\label{t1t2}
  t_1=\log u-\log q-\frac{a}{2}(\log p-\log q)\,,\qquad
  t_2=\log p-\log q,
\end{equation}
and $t_3,\ldots, t_l$ are expressed in terms of $\log q$ and
$\log\zeta$.
This determines the behaviors of $F_0$ and $F_1$ in the limit where
the size of $W_1$ is very large. If we take a further limit where
$q_2\rightarrow 0$ or $p\rightarrow 0$ in addition to $q_1\rightarrow
0$, then it is obvious that $F_0$ and $F_1$ are simply described by
the elliptic polylogarithms of Beilinson and Levin. This is the limit
where $W_2$ decompactifies, namely both the fiber and the base of
$W_2$ are becoming very large.

A short calculation using \eqref{cross1} and \eqref{cross3} shows that
the relation
\begin{equation}\label{Fgmon}
  F_g(u,p,q,\zeta)=F_g(u',q,p,\zeta)\,,
\end{equation}
holds in the limit $q_1\rightarrow 0$, where $g=0,1$ and 
\begin{equation}\label{uprime}
  \log u'=\log u+ \left( - \frac{\tilde \mathcal{I}_2}{12ms} +
      \frac{\mathcal{I}_4}{2m^2s(s+2)} \right) (\log p-\log q)\,.
\end{equation}
It is tempting to conjecture that \eqref{Fgmon} continues to be valid to
all orders in $q_1$ and for all genera $g$.

We can view $\mathcal{F}_0$ as the enumerating function of rational
curves residing entirely in the fibers of $\pi_1:X\rightarrow W_1$.
Let us fix a generic fiber of $\pi_1:X\rightarrow W_1$ and denote it
by $Z$.  Then $\pi:Z\rightarrow B$ with $B\cong \PP^1$ is an elliptic
K3 surface with a section. Here $\pi$ is obtained by restricting
$\pi_2$ to the fiber $f$ of $W_2$. Let us temporarily assume that the
Mordell-Weil group (= the group of sections) of $\pi:Z\rightarrow B$
is trivial so that the section of $\pi:Z\rightarrow B$ is unique.  Let
$S$ and $F$ be respectively the section and a generic fiber of
$\pi:Z\rightarrow B$. Thus we have $S^2=-2$, $F^2=0$ and $S\cdot F=1$.
Besides $S$, rational curves in $Z$ arise from the singular fibers of
$\pi:Z\rightarrow B$.  The singular fibers may be either irreducible
or reducible. In the irreducible cases they must be of Kodaira type
$\mathrm{I}_1$ (a rational curve with a node) or of Kodaira type
$\mathrm{II}$ (a rational curve with a cusp).  A reducible singular
fiber consists of $\PP^1$'s intersecting with one another according to
the pattern of the extended Dynkin diagram of one of the simply-laced
Lie algebras.  Let us assume that we have a single reducible singular
fiber and the rest of the singular fibers are irreducible. The finite
dimensional simply-laced Lie algebra of rank $s_0$ associated with the
reducible fiber is denoted by $\mathfrak{g}_0$.  Let $\tilde
\mathfrak{g}_0$ be the untwisted affine Lie algebra corresponding to
$\mathfrak{g}_0$ and let $\Pi=\{\alpha_0,\ldots,\alpha_{s_0}\}$ be the
set of the simple roots of $\tilde \mathfrak{g}_0$. Thus the $\PP^1$'s
in the reducible fiber correspond to $\Pi$ and their intersection
matrix is equal to the negative of the Cartan matrix of $(\tilde
\mathfrak{g}_0,\Pi)$.  Let $\delta$ and $\Lambda_0$ be as in the
standard theory of affine Lie algebras [Kac]. Thus we have
$(\delta,\delta)=(\Lambda_0,\Lambda_0)=0$ and
$(\delta,\Lambda_0)=1$. $\delta$ is orthogonal to $\Pi$ and
$\Lambda_0$ is orthogonal to $\alpha_1,\ldots,\alpha_{s_0}$, however
$(\Lambda_0,\alpha_0)=1$.  In addition we have the relation
$\alpha_0=\delta-\theta$. We can identify $F$ with $\delta$ and $S$
with $-\Lambda_0-\delta$ with sign flips understood when considering
intersections. When we have a reducible fiber the relative sizes of
the $\PP^1$'s give extra contributions to $\Pic(Z)$.  Thus the
cohomology classes
$[\Lambda_0],[\alpha_0],[\alpha_1],\dots,[\alpha_{s_0}]$ will span
$\Pic(Z)$.

The complexified K{\" a}hler moduli space of $X$ will be given by the
complexified K{\" a}hler cone divided by some discrete group. This
group will arise from the monodromy of the fibrations of $X$.  Since
we have assumed that there are no reducible singular fibers in the
$K3$ fibration $\pi_1:X\rightarrow W_1$, the monodromy-invariant part
of $\Pic(Z)$ and $D_1$ will span $H^2(X,\ZZ)$.  It has been argued
that from the monodromy of $\pi_1:X\rightarrow W_1$ an outer
automorphism of $\mathfrak{g}_0$ can appear [AG,BIKMSV] in addition to
the inner automorphisms described by the Weyl group. In such a case we
divide $\mathfrak{g}_0$ by the outer automorphism and the resulting
simple Lie algebra is denoted by $\mathfrak{g}$. Otherwise we simply
set $\mathfrak{g}=\mathfrak{g}_0$.  Let $s$ be the rank of
$\mathfrak{g}$.  It seems natural to assume that the
monodromy-invariant part of $\Pic(Z)$ has signature $(1,s+1)$ and
contains $Q^\vee(-1)$ where $Q^\vee$ is the coroot lattice of
$\mathfrak{g}$.  We conjecture that $\mathfrak{g}$ coincides with the
one we have been using for the description of (nearly holomorphic)
Weyl-invariant Jacobi forms.  If the Mordell-Weil group of
$\pi:Z\rightarrow B$ is non-trivial or if there are more than one
reducible singular fibers in $\pi:Z\rightarrow B$, the above picture
should be modified appropriately.

What we have conjectured
implies that the complexified K{\" a}hler moduli space contains (a
suitable compactification of) the space
\begin{equation}
  (\HH\times \CC^s)/( SL_2(\ZZ) \ltimes \hat W)\,,
\end{equation}
which is coordinatized by $\tau$ and $z$ as a codimension two
submoduli space.  Here $\hat W=\tilde W \ltimes Q^\vee$ with $\tilde W
=W \ltimes Q^\vee$ being the affine Weyl group of $\mathfrak{g}$. If
we go to a further boundary we will have
\begin{equation}
  \CC^s/\tilde W\,,
\end{equation}
which is coordinatized by $z$.  
The last space also arises from
possible connections to heterotic string theory on $K3\times T^2$.
Let $G$ be a compact simple simply-connected Lie group associated with
$\mathfrak{g}$. We fix a maximal torus $T\subset G$. Let $E$ be an
elliptic curve. By fixing a base point we identify $E$ with the $T^2$
of $K3\times T^2$ on which heterotic string is compactified. The
Wilson lines we can put on $E$ is described by the set of flat
$G$-bundles on $E$ up to isomorphism. There is a natural bijection
[FMW,D] between such set and
\begin{equation}
  \Hom(\pi_1(E),T)/W\cong (E\otimes_\ZZ Q^\vee)/W\,.
\end{equation}
However, apparently we have $(E\otimes_\ZZ Q^\vee)/W\cong \CC^s/\tilde
W$. It is known [L,BS,FMW] that $(E\otimes_\ZZ Q^\vee)/W$ is a
weighted projective space with weights $a_0^\vee, \ldots, a_s^\vee$.

We should also note that the generators $\varphi_0,\ldots,\varphi_s$
of the ring $\oplus_{k,m}J_{k,m}(\mathfrak{g})$ define (part of) the
(inverse) mirror maps (or the solutions to the Jacobi inversion
problems). Actually this was one of the original motivations for the
study of Weyl-invariant Jacobi forms [Sai]. See also [Sat].

Further restrictions on $\phi_{-2,m}$ may arise from the following
consideration.  Let $Z$ be as before an elliptic $K3$ surface with a
section but assume for simplicity that all the singular fibers are of
Kodaira type ${\mathrm I}_1$. Hence there are $24$ of them.  Let $S$
be a section and let $F$ be a generic fiber as before.  As argued in
many works [YZ,Be,Go,BrL], the number of rational curves with $n$
nodes corresponding to the class $S+nF$ is given by $p_{24}(n)$ where
\begin{equation}
  \frac{q}{\Delta(\tau)}=\sum_{n=0}^\infty
  p_{24}(n)q^n=1+24\,q+\cdots\,, \qquad
  \Delta(\tau):=q\prod_{n=1}^\infty(1-q^n)^{24}\,.
\end{equation}
Notice  that
we have a piece
\begin{equation}\label{firsthecke}
   p (\phi_{-2,m}\vert V_1)(\tau,z)=p \phi_{-2,m}(\tau,z)\,,
\end{equation}
in $\mathcal{F}_0$.  Thus it might be tempting to relate
$\phi_{-2,m}(\tau,z)$ to $1/\Delta(\tau)$. However, $1/\Delta(\tau)$
does not have modular weight $-2$. Presumably the source of the
discrepancy partly lies in that we are not considering a single
elliptic $K3$ but a family of elliptic $K3$ surfaces parametrized by
$W_1\cong \PP_1$. Thus the rational curves in $D_1$ will come in
continuous families. For instance, if there are smooth rational curves
in a Calabi-Yau threefold parametrized by a smooth parameter space
$\mathcal{B}$ the virtual number of rational curves is given by
$(-1)^{\dim \mathcal{B}}\chi(\mathcal{B})$ [CdFKM,CFKM,Kat]. Since we
have $X\supset W_2\cong \Sigma_a$ and the parameter associated with
$f\cong \PP^1$ is given by $q_2=pq^{-1}$, we should have
$c(-1,0)=(-1)(2-2\cdot 0)=-2$ and except for this case
$c(-n,\gamma)=0$ $(n \in \NN)$. Thus $-2$ replaces the na\"{ \i}ve
expectation of $1$. (The point $t_2=0$ in the K{\" a}hler moduli space
of $X$ is the place where $f\cong \PP^1$ shrinks. This point is the
place where the U(1) gauge symmetry is enhanced to SU(2) producing two
extra massless vector multiplets explaining the physical meaning of
$-2$.)  By equating the terms proportional to $\zeta(3)$ we should
have
\begin{equation}\label{constant}
  c(0,0)=-\chi\,.
\end{equation}
This tells us that the na\" {\i}ve expectation $24$ should be replaced
by $-\chi$. Actually there should be infinitely many cases where the
virtual number of rational curves are given by $-\chi$ considering the
expression of $\mathcal{F}_0$.  The very appearance of $-\chi$ as the
virtual number of rational curves has been observed and proved for
certain Calabi-Yau threefolds with elliptic fibrations. See \S 4.1 of
[BKK] and [KMV].

We  also note that expressions similar to \eqref{firsthecke} appeared
in different contexts [KMV,HSS].

Taking into account  the above considerations we assume that
$\phi_{-2,m}(\tau,z)$ can  be expressed as
\begin{equation}\label{weylass}
  \phi_{-2,m}(\tau,z) = 
    -\frac{2\,\Phi_{10,m}(\tau,z)}{\Delta(\tau)}\,, 
\end{equation}
where $\Phi_{10,m}(\tau,z)$ belongs to $ J_{10,m}(\mathfrak{g})$ with
$\Phi_{10,m}(\tau,z)=1+O(q)$. Note that since $\Phi_{10,m}(\tau,0)$ is
a modular form of weight ten, we should have
$\Phi_{10,m}(\tau,0)=E_4(\tau)E_6(\tau)$. Of course $\Phi_{10,m}$ can
be expanded (over $\CC$) in the basis of the structure theorem [Wir]
mentioned in \S 2, however we generally expect that the Fourier
coefficients of $\phi_{-2,m}$ are integers.

 With the assumption \eqref{weylass} it is
easy to see that
\begin{equation}
  \mathcal{I}_0=240\,,\quad \tilde \mathcal{I}_0=264\,,\quad \tilde
  \mathcal{I}_2=\mathcal{I}_2\,,
\end{equation}
and
\begin{equation}
  \frac{\mathcal{I}_4}{2m^2s(s+2)}=\frac{\mathcal{I}_2}{12ms}-1\,,
\end{equation}
by \eqref{minus2-2}.  Hence we have simplified expressions for
$\mathcal{F}_0$ and $\mathcal{F}_{0,1}$:
\begin{equation}
  \begin{split}
    \mathcal{F}_0 =\left(\frac{m}{2} -
      \frac{\mathcal{I}_2}{24s}\right) \log p\cdot(\log
    \zeta,\log\zeta)+ \frac{1}{3}(\log
    q)^3-\frac{\mathcal{I}_2}{24s}\log q\cdot(\log\zeta,\log\zeta) \\ 
    +\frac{1}{12}\sum_{\gamma>0}c(0,\gamma)(\gamma,\log \zeta)^3
    +\frac{c(0,0)}{2}\zeta(3) + \sum_{(\ell,n,\gamma)>0} c(\ell
    n,\gamma) \Lic_3(p^\ell q^n\zeta^\gamma)\,,
  \end{split}
\end{equation}
and
\begin{equation}
  \begin{split}
    \mathcal{F}_{0,1}= -12\log
    p -22\log
    q+\inv{2}\sum_{\gamma>0}c(0,\gamma)(\gamma,\log\zeta)\\ 
    +\sum_{(\ell,n,\gamma)>0}\tilde c(\ell n,\gamma)\Lic_1(p^\ell q^n
    \zeta^\gamma)\,.
  \end{split}
\end{equation}
Also \eqref{uprime} is simplified as
\begin{equation}
  \log u'=\log u -(\log p-\log q)\,.
\end{equation}

It is convenient to introduce the classical parts of $F_0$ and $F_1$
arising from constant instanton maps:
\begin{equation}
  F_0^{\rm cl}:=\inv{6}\sum_{i,j,k=1}^l\kappa_{ijk}t_it_jt_k\,,\qquad
  F_1^{\rm cl}:=-\inv{2}\sum_{i=1}^l\rho_it_i\,.
\end{equation}
We have to express $t_i$ in terms of the logarithms of $p$, $q$ and
$\zeta$ so that we can take the limits $q_i\rightarrow 0$ in
$F_0-F_0^{\rm cl}$ and $F_1-F_1^{\rm cl}$. We expect that in addition to
\eqref{t1t2} we should have
\begin{align}
&t_3=\log q-(\gamma_0,\log\zeta)\,,\\
&t_{i+3}=(\Lambda_i,\log\zeta)\,,\quad (i=1,\ldots,s)\,,
\end{align}
where $\Lambda_i$ $(i=1,\ldots,s)$ are the fundamental weights of
$\mathfrak{g}$ and $\gamma_0$ is some positive weight. Then
\eqref{Fgmon} turns into
\begin{equation}
  F_g(q_1,q_2,q_3,q_4,\ldots,q_l) =
  F_g\bigl(q_1{q_2}^{a-2},{q_2}^{-1},q_2q_3,q_4,\ldots,q_l\bigr)\,.
\end{equation}

In the rest of this section we present several conjectural examples as
applications of the general theory developed so far. They are not
particularly new since they have more or less appeared in the
literature. We hope we can report on other examples in the near
future.

Let $X(w_1,\ldots,w_5)_{h^{11},h^{12}}$ symbolically denote a
Calabi-Yau threefold, with the subscripts denoting its Hodge numbers,
obtained from the hypersurface of degree $w_1+\cdots+ w_5$ in the
weighted projective space $\PP(w_1,\ldots,w_5)$.

\bigskip 

{\noindent \bf Example 1.} Take $\mathfrak{g}=\mathsf{A}_1$
and $m=1$. Thus $Q=Q^\vee=\sqrt 2 \ZZ$ and $P=\inv{\sqrt{2}}\ZZ$.  Set
$x=\inv{\sqrt{2}}z$ and $\xi=\mathbf{e}[x]$. Let $E_{k,m}(\tau,x)$ be
the Eisenstein-Jacobi series of weight $k$ and index $m$ as defined in
[EZ].  Consider [Kaw1,CCL]\footnote{This choice is not surprising. Due
  to the structure theorem mentioned in \S 2,  $\Phi_{10,1}$ can always
  be expressed as $\Phi_{10,1}=cE_6E_{4,1}+(1-c)E_4E_{6,1}$ for some
  number $c$.}
\begin{equation}
  \begin{split}
    &\phi_{-2,1}(\tau,z)=-\frac{(12-a)E_6(\tau)E_{4,1}(\tau,x)+
      (12+a)E_4(\tau)E_{6,1}(\tau,x)}{12\Delta(\tau)}\\ &= -
    \frac{2}{q} -2{\xi}^{2}+\left (32+12\,a\right )
    \xi+420-24\,a+{\frac {32+12\,a}{\xi}}-\frac{2}{\xi^2}\\& +
    \Bigl(\left (420-24\,a\right ){\xi}^{2}+\left (53760+96\,a\right )
    \xi+174528-144\, a\\&+{\frac {53760+96\,a}{\xi}}+{\frac
      {420-24\,a}{{\xi}^{2}}} \Bigr)q +O(q^2)\,,
\end{split}
\end{equation}
where $a=0,\ldots,12$.
Since  $\mathcal{I}_2=12+6a$, we have  predictions 
\begin{equation}
  \begin{split}
    F_0^{\rm cl}
&= \log
    u\left (\log p\log q-{(\log \xi)}^{2}\right )- \frac{a}{2}
    \,\log p{(\log \xi)}^{2}\\&\ \ +{\frac {{(\log q)}^{3}}{3}} -
    \left(\frac{a}{2} + 1 \right)\log q (\log \xi)^2 + \left(a +
      \frac{4}{3}\right){(\log \xi)}^{3}\,,
  \end{split}
\end{equation}
and 
\begin{equation}
   F_1^{\rm cl}=-12\log u-12\log p -22\log q
  +(6a+14)\log\xi\,.
\end{equation}
Choosing $\gamma_0=\sqrt{2}$ we set
\begin{equation}
\ t_3=\log q-2\log \xi\,,\qquad  t_4=\log \xi\,. 
\end{equation}
Therefore we find 
\begin{equation}
    \rho_4=156-12 a\,,
\end{equation}
and
\begin{equation}
  \begin{split}
    F_0^{\rm cl}&= t_1\left
      ({t_2}\,{t_3}+{{t_3}}^{2}+2\,{
        t_2}\,{t_4}+4\,{t_3}\,{t_4}+3\,{{t_4}}^{ 2}\right )\\ &\ +
    \frac{a}{2}\,{{t_2}}^{2}{t_3}+\left (1+\frac{a}{2} \right )
    {t_2}\,{{t_3}}^{2}+\frac{4}{3}\,{{t_3 }}^{3}
    +a\,{{t_2}}^{2}{t_4}+2\left (a+2\right ){t_2}\,{
      t_3}\,{t_4}\\&\ +\left (4+a\right )
    {t_2}\,{{t_4}}^{2}+8\,{{t_3}}^{2}{t_4}+\left (14-a\right )
    {t_3}\,{{t_4 }}^{2}+\left (8-a\right ){{t_4}}^{3}\,.
  \end{split}
\end{equation}
Note that this is consistent with $\nu(X,D_1)=1$ and $\nu(X,D_2)=2$.

The would-be Calabi-Yau threefold $X$ must satisfy $\chi=-420+24a$.
For $a=2$ [CCL] and $a=12$ [Kaw1], there are known candidates for $X$.
They are respectively given by $X(10,6,2,1,1)_{4,190}$ and
$X(10,3,3,2,2)_{4,70}$. The conjecture for the case $a=2$ was tested
against the B-model calculation of $F_0$ in [CCL] by using the result
of [BKKM].  The test of the case $a=12$, which was not attempted in
[Kaw1], has now been done up to some orders to find a good agreement
with the B-model calculation of $F_0$.  We thank Hosono for kindly
providing the relevant data of the B-model calculation for
$X(10,3,3,2,2)_{4,70}$.

The $K3$ fibers $D_1$ of $X(10,6,2,1,1)_{4,190}$ and
$X(10,3,3,2,2)_{4,70}$ coincide with the degree $10$ hypersurface in
$\PP(5,3,1,1)$. 

\bigskip
{\noindent \bf Example 2.} Take $\mathfrak{g}=\mathsf{A}_1$ and $m=2$.
Consider
\begin{equation}
  \begin{split}
  &\phi_{-2,2}(\tau,z)=-\frac{5E_6(\tau)E_{4,2}(\tau,x)+7
E_4(\tau)E_{6,2}(\tau,x)}{6\Delta(\tau)}\\
&=
-\frac{2}{q}+96\xi+288+\frac{96}{\xi}+\Bigl(
-2{\xi}^{4}+96\,{\xi}^{3}+10192\,{\xi}^{2}+69280\,\xi+123756
\\&+\frac{69280}{\xi}+
\frac{10192}{\xi^2}+\frac{96}{\xi^3}-\frac{2}{\xi^4}\Bigr) q+O(q^2)\,.
\end{split}
\end{equation}
Since  $\mathcal{I}_2=48$, we should have  
\begin{equation}
  \begin{split}
 F_0^{\rm cl}&=
    \log u\left (\log p\log q-2{(\log \xi)}^{2}\right )-
    2\,\log p\,{(\log \xi)}^{2}\\&\ \ +{\frac {{(\log
          q)}^{3}}{3}} -4\log q (\log
    \xi)^2 +8{(\log \xi)}^{3}\,,
\end{split}
\end{equation}
and 
\begin{equation}
  F_1^{\rm cl}=-12\log u-12\log p -22\log q
  +48\log\xi\,.
\end{equation}
Now assume that $a=2$.  We take $t_3$ and $t_4$ as in the previous
example.  Then we find
\begin{equation}
  \rho_4=88\,,
\end{equation}
and
\begin{equation}
  \begin{split}
 F_0^{\rm cl}&=
t_1\left ({t_2}\,{t_3}+{{t_3}}^{2}+2\,{
t_2}\,{t_4}+4\,{t_3}\,{t_4}+2\,{{t_4}}^{
2}\right )\\
&\ + \,{{t_2}}^{2}{t_3}+2{t_2}\,{{t_3}}^{2}+\frac{4}{3}\,{{t_3
}}^{3}
+2\,{{t_2}}^{2}{t_4}+
8\, {t_2}\,{t_3}\,{t_4}\\
&\ + 4\,{t_2}\,{{t_4}}^{2}+8\,{{t_3}}^{2}{t_4}+ 8\,{t_3}\,{{t_4
}}^{2}+\frac{8}{3}{{t_4}}^{3}\,.
\end{split}
\end{equation}

A candidate Calabi-Yau threefold is $X(8,4,2,1,1)_{4,148}$ whose $K3$
fiber $D_1$ is the degree $8$ hypersurface in $\PP(4,2,1,1)$.
This example should definitely be related to the proposal in [C]. The
nearly holomorphic Weyl-invariant Jacobi form $\phi_{0,2}$ for this
model coincides with the one considered in [GN2].

\bigskip 

{\noindent \bf Example 3.} Take $\mathfrak{g}=\mathsf{E}_8$
and $m=1$. Hence $P=Q=Q^\vee$. We consider
the case $a=12$ and take
\begin{equation}
  \phi_{-2,1}(\tau,z)=-\frac{2E_6(\tau)
    \theta_{0,1}(\tau,z)}{\Delta(\tau)}= -
  \frac{2}{q}+960-2\sum_{(\gamma,\gamma)=2} \zeta^\gamma +O(q)\,,
\end{equation}
where $\theta_{0,1}(\tau,z)=\sum_{\gamma \in
  Q}q^{\frac{(\gamma,\gamma)}{2}}\zeta^\gamma$.  This choice
corresponds to the case studied in [HaM].  Since $\mathcal{I}_2=-480$,
we can predict that
\begin{equation}
    \begin{split}
      F_0^{\rm cl}&=\log u\left (\log p\log q -
        \inv{2}(\log\zeta,\log\zeta)\right )\\&+3 \log p\cdot(\log
      \zeta,\log\zeta)+ \frac{1}{3}(\log q)^3+\frac{5}{2}\log
      q\cdot(\log\zeta,\log\zeta)\\& -
      \frac{1}{6}\sum_{\substack{\gamma>0\\ 
          (\gamma,\gamma)=2}}(\gamma,\log \zeta)^3\,,
\end{split}
\end{equation}
and 
\begin{equation}
  F_1^{\rm cl}=-12\log u-12\log p -22\log
  q -2(\boldsymbol{\rho},\log\zeta)\,,
\end{equation}
where $\boldsymbol{\rho}$ is the Weyl vector of $\mathsf{E}_8$.
Since $P=Q$, we can take [HaM]
\begin{equation}
  t_3=\log q-(\theta,\log \zeta)\,,\qquad t_{i+3}=(\alpha_i,\log
  \zeta)\,, \quad (i=1,\ldots, 8)\,.
\end{equation}
With this choice  we obtain that
\begin{equation}
  (\rho_4,\ldots,\rho_{11})=(368, 548, 732, 1092, 900, 704, 504,
  300)\,.
\end{equation}

A candidate Calabi-Yau threefold is $X(42,28,12,1,1)_{11,491}$ as
found in [HaM].  We remark that there are cousins of this threefold:
$X(30,20,8,1,1)_{10,376}$ and $X(24,16,6,1,1)_{9,321}$. They are
related respectively to $\mathsf{E}_7$ and $\mathsf{E}_6$. It would
be interesting to find out the appropriate  nearly
holomorphic Weyl-invariant Jacobi forms for these Calabi-Yau threefolds.

\section{Discussions}

In this paper we have proposed the asymptotic behaviors of $F_0$ and
$F_1$ for a class of Calabi-Yau threefolds simultaneously admitting a
$K3$ fibration and an elliptic fibration. The key ingredients in this
proposal were the functions $\Lic_1^\phi(p,q,\zeta)$ and
$\Lic_3^\phi(p,q,\zeta)$ which were constructed by liftings. Notably in the
limit where the base of the elliptic fibration decompactifies, $F_0$
and $F_1$ are described by the weighted sums of the elliptic
polylogarithms of Beilinson and Levin.  This should generally be true
if the Calabi-Yau threefold is elliptic regardless of the existence of
a $K3$ fibration structure. String duality is in a sense the search of
a conformal field theory spectrum in some asymptotic region of the
moduli space of one string theory. In this sense elliptic fibrations
seem to be more inherent to string duality phenomena.

What about the asymptotic behavior of $F_g$ $(g\ge 2)$?  It seems
natural to presume that an essential piece of $F_g$ $(g\ge 2)$ in the
limit $q_1\rightarrow 0$ is captured by the {\it automorphic form} 
of weight $2g-2$ [Gr,Bo1]:
\begin{equation}\label{generalF}
  \mathcal{F}_g(p,q,\zeta):= \sum_{\ell=0}^\infty p^\ell
  (\phi_{2g-2,m}\vert V_\ell)(\tau,z)\,,\quad (g\ge 2)\,,
\end{equation}
where $\phi_{2g-2,m}\in J^{\rm nh}_{2g-2,m}(\mathfrak{g})$ will be
constructed from $\phi_{-2,m}$ by appropriate applications of the
differential operators $\mathcal{D}_{k,m}$ and multiplications of
$E_4$ and $E_6$ so that the resulting expression has weight
$2g-2$. Thus $\phi_{2g-2,m}$ will have a single pole at $q=0$ and
\eqref{nondeghecke} tells us that \eqref{generalF} scales as
${t_2}^{2-2g}$ when $t_2 \sim 0$. In general \eqref{generalF} has
poles of order $2g-2$ along rational quadratic divisors
[Bo1,Th9.3]. This is an expected behavior of $F_g$ [BCOV].  Note also
that in contrast to the cases $g=0$ and $g=1$ there do not appear any
powers of logarithms. This is consistent since there is no restriction
on the number of marked points when considering the genus $g$ $(\ge
2)$ Gromov-Witten invariants of Calabi-Yau threefolds.  Using
\eqref{negativeinv} it is easy to see that $\mathcal{F}_g(p,q,\zeta)=
\mathcal{F}_g(q,p,\zeta)$ for $g\ge 2$ which is part of the
automorphic properties of $\mathcal{F}_g(p,q,\zeta)$ and is consistent
with the conjecture \eqref{Fgmon}.  However, we do not know how
precisely $\mathcal{F}_g$ is related to $F_g$ at the moment.  More
detailed properties of $\mathcal{F}_g$ are currently under study.

A most urgent problem would be to explain the appearance of the
nearly holomorphic Weyl-invariant Jacobi form purely from the geometry
of the Calabi-Yau threefold $X$. A modest starting point would be to
establish the part \eqref{firsthecke} in $F_0$.

We studied $\Lic_r^\phi(p,q,\zeta)$ only for $r=1$ and $r=3$ since these
are of direct relevance to the main object of this paper. However, the
procedure should be general and in principle one may consider
$\Lic_r^\phi(p,q,\zeta)$ for larger values of $r$. This may be of
independent interest though the connection to string theory is not so
clear.

\section*{Acknowledgments} The author would like to
express his sincere gratitude to the Taniguchi foundation for generous
support and the organizers of the symposium for their kind invitation.
He is grateful to K. Saito, M-H. Saito, Y. Shimizu for discussions on
the subject.  He also thanks S. Hosono for providing a relevant
B-model calculation.

\end{document}